\documentclass[12pt]{article}

\usepackage{graphicx,,color}
\usepackage{amsmath,amssymb,amsfonts}
\usepackage{array}
\usepackage{cite}
\usepackage{rotating}

\newcommand{\lsim}{\raisebox{-0.13cm}{~\shortstack{$<$ \\[-0.07cm] $\sim$}}~} 
\newcommand{\gsim}{\raisebox{-0.13cm}{~\shortstack{$>$ \\[-0.07cm] $\sim$}}~} 
\newcommand{\beq}{\begin{eqnarray}} 
\newcommand{\eeq}{\end{eqnarray}} 
\newcommand{\tb}{\tan\beta}

\begin{document}

\begin{flushright}LPT-Orsay--16--14 \end{flushright} 

\vspace*{3mm}

\begin{center}

{\large\bf Threshold enhancement of diphoton resonances}

\vspace{7mm}

{\sc Aoife~Bharucha$^{1}$, Abdelhak Djouadi$^{2,3}$,  Andreas Goudelis$^{4}$ }

\vspace*{7mm} 

{\small 

$^1$ Centre de Physique Th\'eorique,  CNRS \& Universit\'e Aix Marseille \& 
Universit\'e Toulon, \\ UMR 7332, F-13288  Marseille, France.\\[1mm]
$^2$ Laboratoire de Physique Th\'{e}orique, CNRS and Universit\'{e} Paris-Sud, \\
B\^{a}t. 210, F–91405 Orsay Cedex, France. \\[1mm]
$^3$ Theory Department, CERN, CH 1211 Geneva 23, Switzerland.\\[1mm]
$^4$ Institute of High Energy Physics, Austrian Academy of Sciences, \\ 
Nikolsdorfergasse 18,  1050 Vienna, Austria.

}

\end{center}

\vspace*{2mm}

\begin{abstract}
The data collected by the LHC collaborations at an energy of 13 TeV indicates the presence of an excess in the diphoton spectrum that would correspond to a resonance of a 750 GeV mass. The apparently large production cross section  is nevertheless very difficult to explain in minimal models. We consider the possibility that the resonance is a pseudoscalar boson $A$ with a two--photon decay mediated by a charged and uncolored fermion having a mass at the $\frac12 M_A$ threshold and a very small decay width, $\ll 1$ MeV; one can then generate a large enhancement of the $A\gamma\gamma$ amplitude which explains the excess without invoking a large multiplicity of particles propagating in the loop, large electric charges and/or very strong Yukawa couplings. The implications of such a threshold enhancement are discussed in two explicit scenarios: i) the Minimal Supersymmetric Standard Model in which the $A$ state is produced via the top quark mediated  gluon fusion process  and decays into photons predominantly through loops of charginos with masses close to $\frac12 M_A$ and ii) a two Higgs doublet model in which $A$ is again produced by gluon fusion but decays into photons through loops of vector--like charged heavy leptons. We also comment on a minimal scenario in which the $A$ state couples only to photons through a heavy lepton loop and is both produced and decays through this coupling.  In all these scenarios, while the mass of the charged fermion has to be adjusted to be extremely close to half of the $A$ resonance mass, the small total widths are naturally obtained if only suppressed three-body decay channels occur. Finally,  the implications of some of these scenarios for dark matter are discussed.
\end{abstract}

\newpage


\subsection*{1. Introduction} 

There is presently considerable excitement in the particle physics community as the  ATLAS and CMS collaborations have reported an excess in the data collected from LHC collisions at an energy of 13~TeV, corresponding to a possible resonance with a mass of 750 GeV decaying into two photons \cite{diphoton}. Assuming the resonance to be a scalar boson denoted $\Phi$, the production cross section times the decay branching ratio for the final state  $pp\! \to\! \Phi\!  \to \! \gamma \gamma$ is very large, $=\! 6\pm 2$~fb~\cite{cite-few}.  Such a rate is very difficult to accommodate in minimal versions of theories that are often considered to be attractive extensions of the Standard Model (SM). For example, it has been shown \cite{ADM} that in its Minimal Supersymmetric extension (MSSM) \cite{Reviews}, while there are additional Higgs bosons beyond the already observed one that can indeed be identified with the 750 GeV state, the diphoton rate cannot be generated using purely the MSSM particle content. Hence, the $\Phi$ resonance must be accompanied by additional massive charged particles to enhance the $\Phi \gamma\gamma$ decay amplitude and, eventually, the $\Phi gg$ amplitude in the likely case where the resonance is produced via the gluon fusion mechanism, $gg\to \Phi$. 

An interesting possibility would be that these additional particles are electrically charged and non--colored (generally vector--like) fermions that contribute only to the $\Phi \to \gamma\gamma$ decay; see e.g. Ref.~\cite{ADM}. However, in this specific case, the large enhancement of the $\Phi \gamma\gamma$ amplitude would require either $i)$ several charged fermions, and/or $ii)$  large electric charges, and/or $iii)$ strong Yukawa couplings. All these requirements could unfortunately put perturbation theory under jeopardy~\cite{PT}. The same problem occurs, although to a lesser extent, in certain scenarios where additional colored particles contribute to the $\Phi gg$ vertex \cite{cite-few,ADM,PT}. 

One means by which this problem could be alleviated would be to assume that $\Phi$ is a pseudoscalar state $\Phi \equiv A$ and that the charged fermions running in the $A\gamma\gamma$ loop have masses near the $m = \frac12 M_A$ kinematical threshold \cite{ADM}. In this case, the form factor $A_{1/2}^\Phi$~\cite{Reviews,venerable} that characterizes the loop contributions of spin--$\frac12$ fermions as functions of the scalar to fermion mass ratio (and which depends on the parity of the spin-zero state) becomes maximal and much larger than in the very heavy or very light fermion mass limits.  Nevertheless, even in this particular case, the obtained $A_{1/2}^A$ form factor is not sufficient to explain the large diphoton rate in a minimal way and without endangering perturbation theory. We should note that in the context of the MSSM, even for masses close to $M_A$, the contribution of the two $\chi_1^\pm$ chargino states to the $A \to \gamma \gamma$ rate  is too small as their couplings are very weak. 

In this paper, still assuming a pseudoscalar resonance and a charged and uncolored fermion  with a mass close to the $\frac12 M_A$ threshold, we invoke an additional mechanism to enhance the $A\gamma\gamma$ loop amplitude: the charged fermions will form S--wave (quasi) bound states resulting in a Coulomb singularity developing very close to this kinematical threshold \cite{singular,MSY,GreensFn}. This singularity is regulated by  the total decay width of the charged fermion which, if it is very small, say $\Gamma \lsim 1$ MeV, allows an enhancement of the $A\gamma\gamma$ amplitude by a large factor. This means that with only one singly charged fermion having a reasonable Yukawa coupling to the resonance, one could generate a $A\to \gamma\gamma$ amplitude that is sufficiently large to accommodate the LHC diphoton signal. This interesting possibility will be studied in two specific examples.  

We first reconsider the MSSM \cite{ADM}, assuming that the CP--odd $A$ state corresponds to the 750 GeV  diphoton resonance and has a strong top quark Yukawa coupling. This already allows for a significant cross section in the top induced  $gg\to A$ process and the large total width, $\Gamma_A \approx 45$ GeV,  hinted by the ATLAS data \cite{diphoton}. The required $A \to \gamma\gamma$ decay rate is then generated by loops of charginos with a mass $m_{\chi_1^\pm}=\frac12 M_A$ for which the $A\gamma \gamma$ form--factor develops a Coulomb singularity that is regulated by a chargino width $\Gamma_{\chi_1^\pm} \ll 1$ MeV. Such a small decay width can be achieved naturally by imposing that the only possible decay mode, the one into the stable lightest neutralino and a fermion pair, occurs only at the three--body level and is strongly suppressed. One can then have a large threshold enhancement which easily explains the LHC diphoton data in this minimal supersymmetric model. 

In a second scenario, we consider either the MSSM or a two Higgs doublet model (2HDM) \cite{2HDM} in which the $A$ state still corresponds to the new resonance as discussed above but where the required $A \to \gamma\gamma$ decay rate is now generated by two vector--like doublets and singlets of heavy leptons \cite{ADM}. The lightest charged lepton $E$ has a mass very close to threshold $m_{E} = \frac12 M_A$ and the Coulomb singularity is again regulated by a small decay width that is obtained by assuming a lighter (possibly stable) neutral lepton $N$ with a mass $m_{E}<m_{N}+M_W$ such that the only available decay mode is the suppressed three--body channel $E \to N W^* \to Nf\bar f'$.  We then explore the regions in the parameters $\Gamma_E$ and $m_E- m_N$ that allow us to obtain the enhancement factor which explains the $\approx 6$ fb $gg\to A \to \gamma\gamma$ production rate at the LHC.

We also comment on a minimal scenario in which the 750 GeV resonance couples only to photons and is therefore both produced and decays through the $A\gamma\gamma$ loop induced coupling \cite{photon}. In order to obtain the LHC diphoton rate in the process $\sigma(pp \to \gamma\gamma \to A \to \gamma \gamma)$, an extremely large two--photon decay width is needed in this case. We show that, again, the additional loop contribution of only one singly charged lepton with mass $m_{E} = 375$ GeV and total width $\Gamma_{E} \approx 1$ keV allows for a sufficient  threshold enhancement of the $A\gamma\gamma$ form--factor to explain the data in this minima scenario.

Finally, in both the MSSM with a stable lightest neutralino and in a 2HDM where the charged lepton $E$ is accompanied by a stable neutral one $N$, we consider the tantalizing possibility that the neutral particles are viable dark matter candidates \cite{DM}. We determine the range of masses and couplings that would allow such a possibility, once the relevant experimental constraints from direct and indirect dark matter searches are imposed.


\subsection*{2. Threshold enhancement of the diphoton width} 

Let us begin by discussing the possibility of a threshold enhancement in the general context 
of a spin--zero CP--even $H$ or CP--odd $A$ state with two--photon decays induced by loops of fermions with color number $N_f^c$, electric charge $e_f$ and  Yukawa couplings $\lambda_{\Phi ff}$ when normalised to their SM-like values, $\lambda_{\Phi ff}^{\rm SM} \!= \! \sqrt 2 m_f/v$ with $v\! =\! 246$ GeV. The two--photon partial decay widths read \cite{venerable,Reviews}
\begin{eqnarray}\label{eq:phigg}
\Gamma(\Phi  \to \gamma\gamma) =  \frac{G_\mu\alpha^2 M_\Phi^3} {128\sqrt{2}\pi^3} 
\bigg| \sum_f N^c_f e_f^2 \lambda_{\Phi ff} A_{1/2}^\Phi (\tau_f) \bigg|^2
\end{eqnarray}
where $\alpha$ is the QED fine structure constant, $\alpha= e^2/4\pi \approx 1/128$ at a scale $M_\Phi$ and $G_\mu$ is the Fermi constant. The form factors $A_{1/2}^\Phi (\tau_f)$ which depend on the variable $\tau_f=M_\Phi^2/4m_f^2$ are given, for the scalar and the pseudoscalar cases, by 
\begin{eqnarray}
& A_{1/2}^{H}  = 2 \left[  \tau_{f} +( \tau_{f} -1) f(\tau_{f})\right]  \tau_{f}^{-2} \, , \ \  
A_{1/2}^{A}  = 2 \tau_{f}^{-1} f(\tau_{f}) \, ,  \label{eq:Af} \\ 
& f(\tau)=\left\{ \begin{array}{ll}  \displaystyle
\arcsin^2\sqrt{\tau } & {\rm for} \; \tau \leq 1 \, , \\
\displaystyle -\frac{1}{4}\left[ \log\frac{1+\sqrt{1-\tau^{-1} }}
{1-\sqrt{1-\tau^{-1}}}-i\pi \right]^2 \hspace{0.5cm} & {\rm for} \; \tau > 1 \, .
\end{array} \right.
\label{eq:formfactors}
\end{eqnarray}
The amplitudes are real for $\Phi$ masses below the $M_\Phi \!= \! 2m_f$ kinematical threshold and develop an imaginary part above. When the loop fermion is much heavier than the $\Phi$ state, $m_f \to \infty$, one obtains $A^{H}_{1/2}\! =\! \! \frac43$ and $A^A_{1/2}\!=\!2$ , while in the opposite limit, $m_f \to 0$, one has $A_{1/2}^{\Phi} \to 0$. 

The maximal values of the form factors are attained near the mass threshold  $m_f = \frac12 M_\Phi$ where one has:  Re($A^H_{1/2}) \approx 2 $ and Re($A^A_{1/2}) \approx \frac12 \pi^2 \approx 5$ for the real parts and Im($A^{\Phi}_{1/2}) \! \approx \! 0$. 
Hence, near threshold, the form--factor $A^\Phi_{1/2}$ is much larger for a CP--odd state and we will thus concentrate on this case in the rest of the discussion.  Furthermore, we will only consider the case where color--neutral fermions contribute in the loops: heavy quarks would also contribute to the $Agg$ loop--induced coupling\footnote{In principle, one expects these quarks to be rather heavy from LHC direct searches, $m_Q \gsim 800$ GeV ~\cite{LHC-VLQ} and hence beyond the $m_Q =\frac12 M_\Phi$ threshold. Nevertheless such a configuration could be possible in some special cases where bound states can form; see e.g.  Ref.~\cite{Q-boundstates}. In the case where the resonance also couples to top quarks, there would also be a significant enhancement of the $A\gamma\gamma$ amplitude near the $M_\Phi \approx 2m_t$ threshold~\cite{MSY,SDGZ}, but it is negligible in practice since a 750 GeV resonance is far from this configuration.} and would generate unacceptably large production rates in the situations which will be considered here (like in the $pp\to t \bar t$ process for instance~\cite{Htt}). 

Nevertheless, the expressions eqs.~(\ref{eq:Af})--(\ref{eq:formfactors}) above do not accurately describe the threshold region $m_f \approx \frac12 M_\Phi$ for the $A\gamma\gamma$ form factor. Indeed, for fermion masses just above but very close to threshold, a Coulomb singularity develops due to the fermions forming S--wave (quasi) bound states~\cite{Coulomb}. This can be taken into account, in a non-relativistic approach, by re-writing the form factor close to threshold as~\cite{MSY}
\begin{equation}
A_{1/2}^A =a+b \times G(0,0;E_f +i \Gamma_f),
\end{equation}
where, to leading order, one  has $a=\frac12 \pi^2$ and $b= 8 \pi^2/m_f^2$ for the real and imaginary parts, $E_f =M_A-2 m_f$ for the distance from the threshold region  and $\Gamma_f$ is the total decay width of the fermion $f$ running in the loop. Here $a$ and $b$ are the perturbatively calculable coefficients obtained from matching the non-relativistic theory to the full theory. $ G(0,0;E_f)$ is the S--wave Green's function of the non-relativistic Schr\"{o}dinger equation in the presence of a Coulomb potential $V(r)=-\alpha/r$. 
The fermion decay width $\Gamma_f$ is introduced in order to regulate the Coulomb singularity in the Green's function with real and imaginary parts~\cite{GreensFn}
\begin{align}
{\rm{Re}}\,G(0,0;E_f+i\Gamma_f)=&-\frac{m_f \,p_-}{4\pi}+\frac{m_f \,p_0}{4\pi}\log\frac{m_f^2 D^2}{p_+^2+p_-^2}+\frac{m_f\, p_0^2}{2\pi}\sum_{n=1}^{\infty}\frac{p_--p_n}{n^2[(p_--p_n)^2+p_+^2]} \, ,\\
{\rm{Im}}\,G(0,0;E_f+i\Gamma_f)=&-\frac{m_f \,p_+}{4\pi}+\frac{m_f \,p_0}{2\pi}\arctan\frac{p_+}{p_-}+\frac{m_f \,p_0^2}{2\pi}\sum_{n=1}^{\infty}\frac{p_+}{n^2[(p_--p_n)^2+p_+^2]} \, ,
\end{align}
where $p_{\pm}=(\frac12 m_f(\sqrt{E_f^2+\Gamma_f^2}\pm E_f))^{1/2}$, $p_n=p_0/n$, $p_0=\frac12 m_f  \alpha$ and $D$ is a renormalization constant which we set to unity in the following as this will only affect our results at higher orders\footnote{In principle, one could calculate higher order corrections to the coefficients $a$ and $b$; however, these are not needed for this preliminary study as, in particular for the QED case, they should be highly suppressed.}~\cite{MSY}. The three terms in the above expressions correspond to the lowest order contribution, a single Coulombic photon exchange and a sum over contributions involving the exchange of $n+1$ Coulombic photons. The position of the first pole in $E_f$ can be obtained by inspecting the denominator of the $n=1$ contribution to the last terms of the equations above. Although the sum in $n$ runs from 1 to $\infty$, the sum converges rather quickly and, in reality, it is sufficient for our purposes to calculate up to $n=100$. 

One should note that while the pole is present in both the real and imaginary parts, the numerator $p_+$ for the imaginary part is suppressed compared to that of the real part $p_- - p_0/n$ in the vicinity of the pole. The large enhancement of the two--photon form factor is  therefore obtained from the real part of the Green's function.

As mentioned earlier, the bound-state formation results in poles in the form factor at energies just below the threshold for pair production of the fermions, regulated by the width of the fermions. The position of these poles in terms of the energy $E_f$  as well as the size of the enhancement therefore depends on the size of the total decay width and the coupling to the photon. It occurs in an extremely narrow range of the mass difference $E_f$, ${\cal O}(0.1$ MeV). 

Besides the fact that the dominant contribution to the enhancement is from the real and not the imaginary part of the form factor, one should note that in the CP--even scalar case, the $H\gamma\gamma$ form factor is P--wave and highly suppressed at the threshold;  therefore, the bound state formation can be neglected as in this case the possible enhancement is negligible. 

\begin{figure}[btp]
\begin{center}
\includegraphics[scale=0.85]{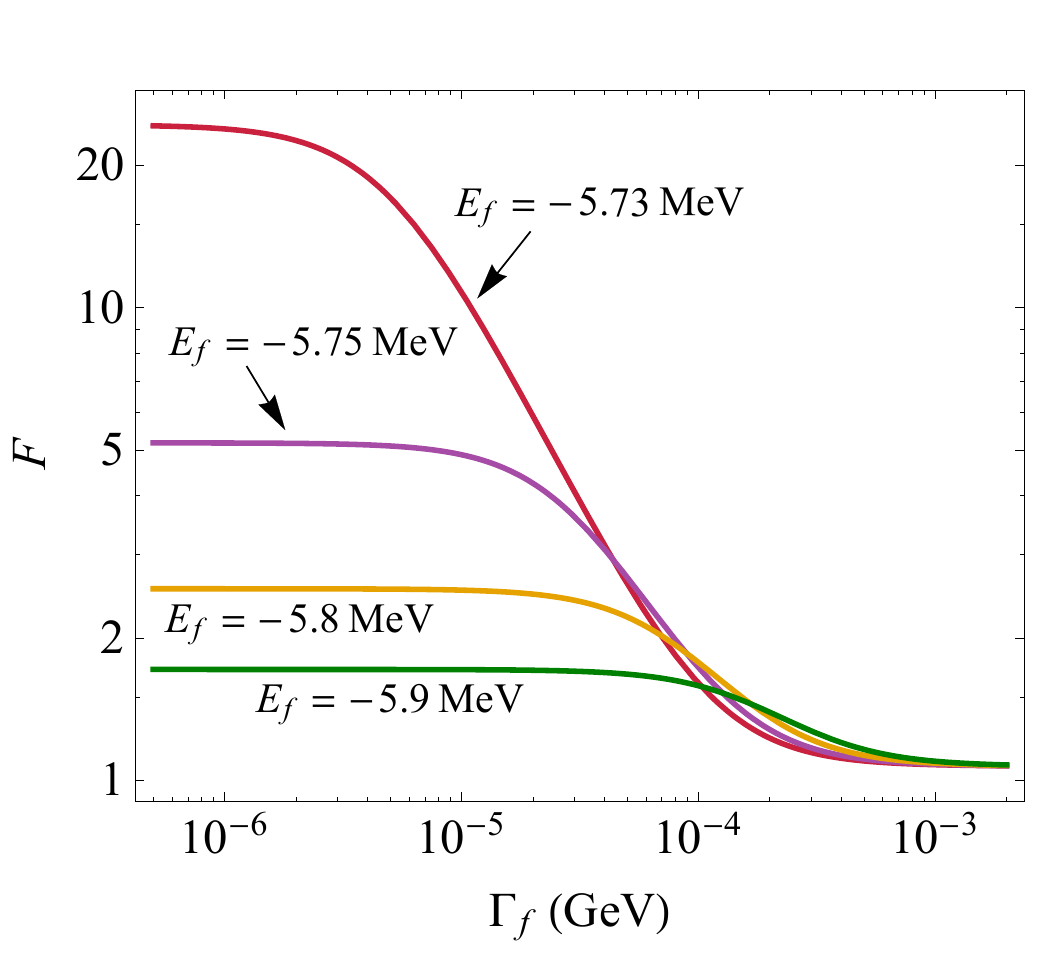}
\vspace*{-2mm}
\caption{The enhancement factor $F$ of the absolute value of the form factor $A_{1/2}^A$ on including the threshold corrections compared to the leading order term, as a function of the fermion total width $\Gamma_f$ for several values of the mass difference $E_f \! =\! 
M_A -2m_f$ as indicated.} 
\label{fig:Ratio-Gamma}
\end{center}
\vspace*{-6mm}
\end{figure}

In Fig.~\ref{fig:Ratio-Gamma}, we display the absolute value of the enhancement factor $F$ defined as
\begin{eqnarray}
F={A^A_{1/2}(\mbox{threshold enhanced})}/{A^A_{1/2}(\mbox{perturbative})},
\end{eqnarray}
 as a function of the  total width of the fermion $\Gamma_f$, for a resonance mass $M_A=750$ GeV and $E_f \!= \! M_A -2 m_f$ values of $\! -\! 5.73,\,-5.75,\,-5.8$ and $-5.9$ MeV. As can be seen, for $\Gamma_f <2$ keV, one can obtain enhancement factors of 20 and beyond. As we shall see in the following section, such a factor $F \approx 20$ could explain the LHC results in the MSSM case,  whereas in our 2HDM scenario a factor $F \approx 5$ would be sufficient, requiring $\Gamma_f\lesssim 50$ keV. The enhancement factor is of course dependent on the choice of $E_f$ or conversely $m_f$, i.e.~whether or not $E_f$ corresponds exactly to the position of the pole.
This is the Achilles heel of our scenario as some ``fine-tuning" is thus necessary.

\subsection*{3. Implications for diphoton resonance scenarii} 

Let us now discuss the implications of this possible threshold enhancement in some scenarii for the 750 GeV $\Phi=A$ resonance, starting with the plain MSSM scenario.  


In the MSSM, two Higgs doublets $\Phi_u$ and $\Phi_d$ are required to break the electroweak symmetry leading to five physical states, two CP--even $h$ and $H$, a CP--odd $A$ and two charged $H^\pm$ bosons \cite{Reviews}. 
In the so--called decoupling limit, $M_A \gg M_Z$, the lighter $h$ state is the Higgs boson observed at the LHC in 2012 and subsequently determined to have  SM--like properties, while 
the $\Phi$ resonance is a superposition of the heavier neutral CP--even $H$ and CP--odd $A$ that are nearly degenerate in mass (as is also the case of the charged Higgs boson). The two states have zero tree--level couplings to vector bosons and similar  couplings  to fermions. The latter are controlled by the ratio of vacuum expectation values $\tan\beta \! = \! v_u/v_d$ which is the only relevant parameter in the Higgs sector of the model besides $M_A \approx M_H$.  For $\tan\beta \approx 1$, the only important Yukawa coupling is the one of the top quark, $y_t = \sqrt 2 m_t/ (v \tan\beta ) \approx 1$. 

At the LHC, the $H/A$ states are mainly produced in the $gg\to \Phi$ fusion mechanism that is mediated by a top quark loop with an amplitude that is given by an expression similar to that of eqs.(\ref{eq:phigg})--(\ref{eq:formfactors}) except for some color factors and the replacement of $\alpha$ with $\alpha_s$ \cite{venerable}. The cross sections are such that  $\sigma (gg \to A) \! \approx \! 2 \sigma(gg\to H)$ and for $M_\Phi \! \approx \! 750$ GeV and $\tb \! \approx \! 1$, one obtains $\sigma(A\!+\!H) \! \approx \! 2$ pb at the $\sqrt s \!=\! 13$ TeV LHC \cite{hMSSM}.  The $\Phi=H/A$ states will then mainly decay into top quark pairs with partial ($\approx$ total) widths that are of order $\Gamma_\Phi \approx  35$ GeV close to the resonance width of 45 GeV favored by ATLAS. Concentrating on the pseudoscalar $A$ resonance,  if the two-photon  decay is generated by the top quark loop only, the  branching ratio for the relevant inputs is BR$(A \to \gamma\gamma) \approx 7 \times  10^{-6}$ \cite{hdecay}. One thus has a resonance production times decay rate of about $  \sigma (gg\to A) \times {\rm BR}(A \to \gamma\gamma) \approx  10^{-2}  \; {\rm fb}$.

In Ref.~\cite{ADM}, the possible loop contributions of the supersymmetric particles have been analyzed and found to be far too small to explain the large two--photon decay rate. Here, we will reconsider the chargino loop contribution to $A \to \gamma\gamma$ in light of the possible threshold enhancement discussed above. These contributions are briefly summarized below. 

The general chargino mass matrix, in terms of the wino and higgsino mass parameters $M_2$ and $\mu$ in the limit $\tan\beta \approx 1$ in  which we specialize,  is given by 
\begin{eqnarray}
{\cal M}_C = \left[ \begin{array}{cc} M_2 & \sqrt{2}M_W \sin \beta
\\ \sqrt{2}M_W \cos \beta & \mu \end{array} \right]
\stackrel{\small \tan\beta=1} \to 
 \left[ \begin{array}{cc} M_2 & M_W  \\ M_W & \mu \end{array} \right]
\label{matrix:ino}
\end{eqnarray}
The two physical chargino states $\chi_1^\pm, \chi_2^\pm$ and their masses are determined through unitary matrices $U$ and $V$ defined and given by ($\sigma_3$ is the Pauli matrix with diagonal values 
$+1, -1$)  
\beq
U^* {\cal M}_C V^{-1}~:~ V= {\cal O}_+ \ , \ U= \bigg\{   
\begin{array}{cc} {\cal O}_- & {\rm if~det~}{\cal M_C} >0  \\ 
\sigma_3  {\cal O}_- & {\rm if~det~}{\cal M_C} <0   \end{array}  \ , \ 
{\cal O}_\pm = \left[\begin{array}{cc} \cos\theta_\pm  & \sin \theta_\pm \\
-  \sin \theta_\pm &  \cos \theta_\pm \end{array} \right]
\eeq
The coupling of the pseudoscalar $A$ boson  to pairs of the  same chargino is given by 
\beq
g_{A \chi^-_i \chi^+_i } =  - \frac{e}{\sqrt 2 \sin\theta_W} \left[\sin\beta V_{i1}U_{i2} + \cos\beta 
V_{i2}U_{i1} \right] \stackrel{\small \tan\beta=1} \to - \frac{e}{2 \sin\theta_W} \left[V_{i1}U_{i2} + 
V_{i2}U_{i1} \right] \label{cp:inos2}
\label{cp:ino}
\eeq
As can be seen, this coupling is maximal for equal admixtures of higgsinos and winos, $|\mu|\approx M_2$. Taking the limit $|\mu|=M_2 \gg M_W$ for simplicity and still $\tan\beta=1$ (a choice which maximizes the $A$ production cross section), the matrix eq.~(\ref{matrix:ino}) is easy to diagonalise. For $\mu \geq 0$,  one obtains for the masses of the charginos and their couplings to the $A$ state 
\beq
\mu > 0 \, :~ m_{\chi_{1,2}^\pm} \simeq \mu \mp M_W \ , \ 
g_{A \chi^-_1 \chi^+_1 } \simeq -g_{A \chi^-_2 \chi^+_2 } \simeq e/(4 \sin\theta_W)
\label{eq:mu+}
\eeq
For large $\mu$ values, the two  states ${\chi_{1,2}^\pm}$ have thus masses that are close to each other and couplings to $A$ of opposite sign; their loop contributions to the $A\gamma\gamma$ amplitude will therefore interfere destructively with each other reducing the $\chi^\pm$ impact
in $A\to \gamma \gamma$ decays. 

\begin{figure}[t]
\begin{center}
\vspace*{-.1cm}
\includegraphics[width=0.33\textwidth]{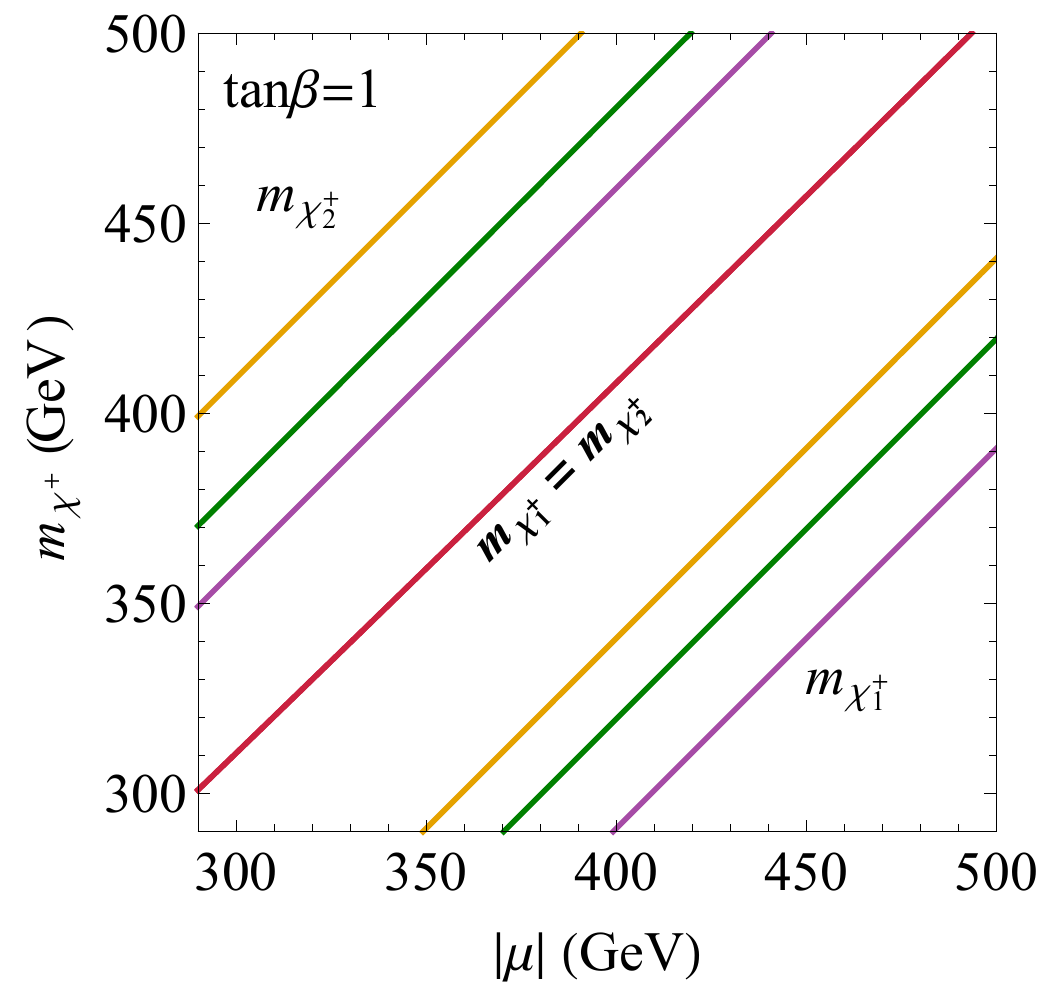}
\includegraphics[width=0.32\textwidth]{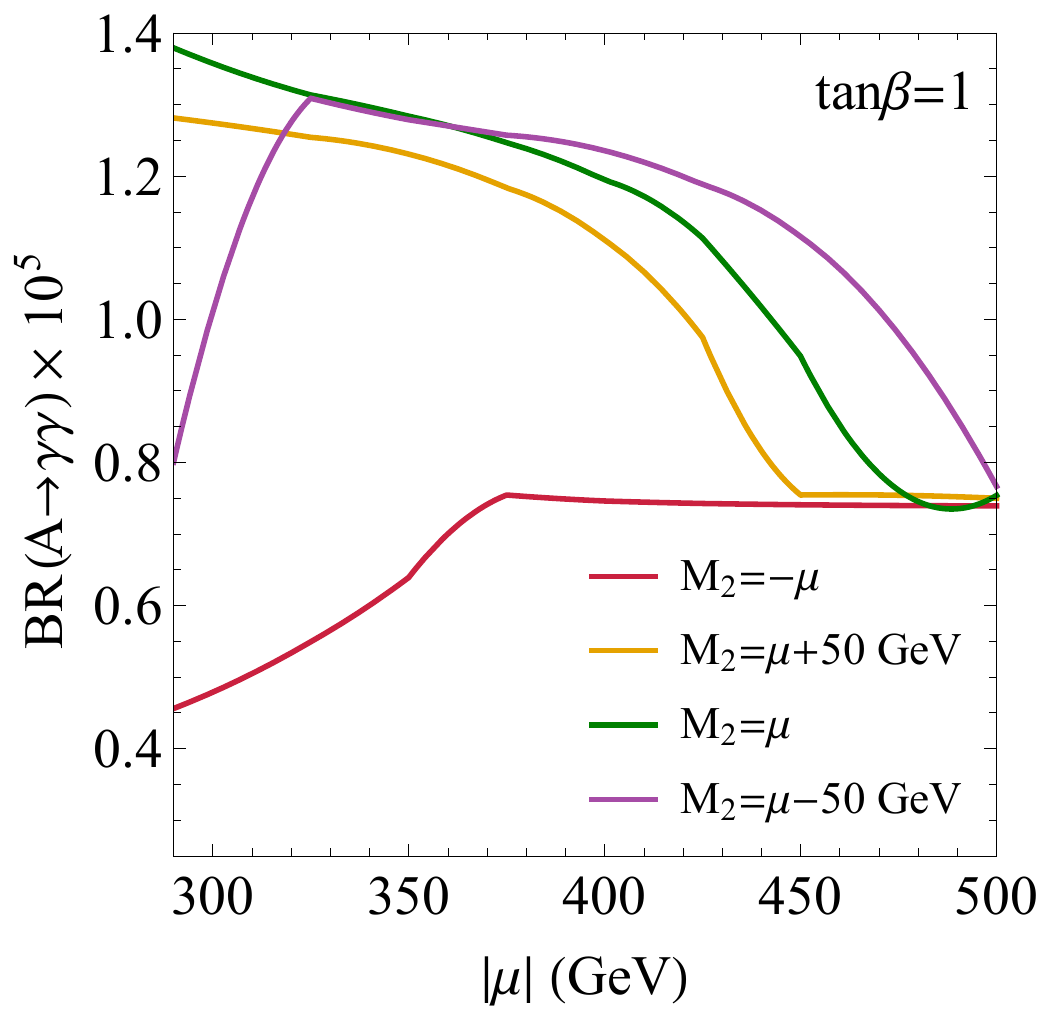}
\includegraphics[width=0.32\textwidth]{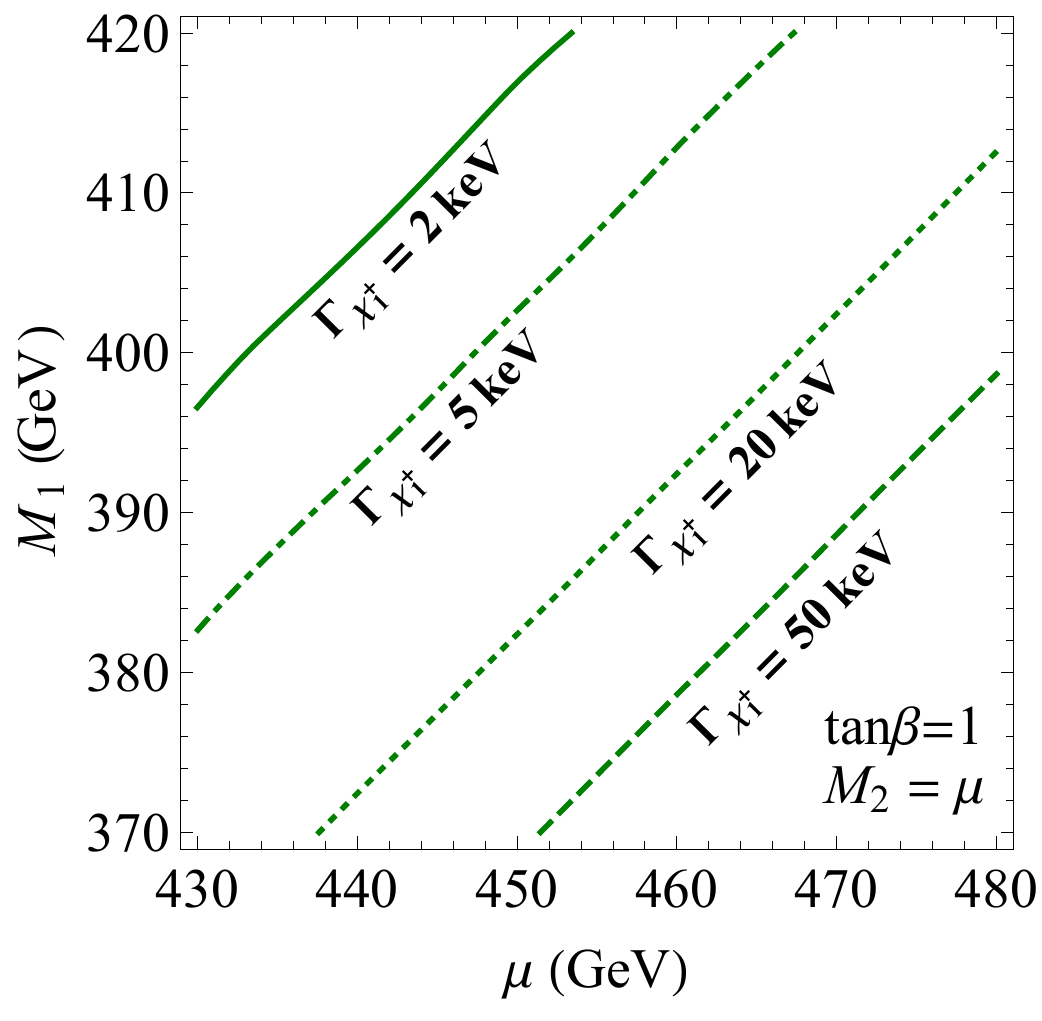}
\vspace*{-.2cm}
\caption{The two chargino masses (left) and the branching  ratio  BR($A\to \gamma\gamma)$ when the chargino contributions are included (centre) as a function of $\mu$ for the values of $M_2$, $\mu , \mu +50$ GeV and $\mu+50$ GeV (for $\mu>0$) and $M_2 = -\mu$ (for $\mu<0$). Contours in the $[\mu, M_1]$ plane for which we obtain a total chargino width of $\Gamma (\chi_1^\pm \to  \chi_1^0 f\bar f')=2,5,20$ and 50 keV (right). The MSSM with $\tan\beta=1$ and $M_A=750$ GeV (and heavy sfermions) is assumed in all cases. } 
\label{fig:inos}
\end{center}
\vspace*{-.4cm}
\end{figure}

In turn, for $\mu <0$, one obtains for the masses and couplings in the limit $|\mu| \approx 
M_2 \gg M_W$, 
\beq
\mu < 0 \, :~ m_{\chi_{1,2}^\pm} \simeq |\mu| 
\ , \ g_{A  \chi^-_1 \chi^+_1} \simeq - g_{A \chi^-_2 \chi^+_2} = {\cal O}(M_W^2/ \mu^2) 
\label{eq:mu-}
\eeq
Hence, the two charginos are nearly degenerate in mass and have suppressed couplings to the
$A$ state. Some numerical examples~\cite{hdecay} for the two options of $\rm{sign}(\mu)$ are shown in Fig.~\ref{fig:inos} where, in the left--hand side, we display the chargino masses for the four possibilities $\mu=M_2, M_2 \pm 50$ GeV when $\mu \geq 0$ and $\mu=M_2$ for $\mu < 0$. Indeed, the masses behave as described in eqs.~(\ref{eq:mu+}) and (\ref{eq:mu-}) for the $\mu>0$ and $\mu<0$ cases respectively. In the central  frame, we show the branching ratio BR($A\to \gamma\gamma)$ when the (un-enhanced) chargino loop contributions are included. As can be seen, the contributions are not the largest for $m_{\chi_1^\pm}=375$ GeV, as one would na\"ively expect since the form factor $A_{1/2}^A \approx \frac12 \pi^2$ is maximal. This effect is due to the negative interference of the two chargino loops  for $\mu >0$ and the small Higgs couplings to charginos in the $\mu <0$ case.

We now focus on the the case $\mu=M_2$ with $\mu >0$ and describe the spectrum for $\tan\beta=1$ when the lightest chargino is at the $m_{\chi_1^\pm} =\frac12 M_A = 375$ GeV threshold (which, here,  occurs for $\mu=455$ GeV). We would have $m_{\chi_2^0} \approx  m_{\chi_1^\pm} \approx \mu - M_W$, $m_{\chi_3^0} \approx \mu$ and $m_{\chi_4^0} \approx \mu+ M_W$ and the lightest neutralino mass can then be chosen via the remaining input that enters the chargino--neutralino sector: the bino mass parameter $M_1$. Here, a careful choice ensures that the total width for the chargino $\chi_1^\pm$ is small, providing the threshold enhancement factor of the $A \chi_1^\pm \chi_1^\mp$ loop form--factor needed to explain the LHC data. Indeed, in the R--parity conserving scenario that we consider here, the only possible decay of the chargino $\chi_1^\pm$ will be into  the lightest neutralino $\chi_1^0$ (which is stable) and a $W$ boson. If the mass splitting $m_{\chi_1^\pm} -m_{\chi_1^0}$ is small, the $W$ boson is off--shell and decays into light fermions through the three--body decay  $\chi_1^\pm \to \chi_1^0 W^* \to  \chi_1^0 f\bar f'$, which has a very small partial (= total) width. 

In the right panel of Fig.~\ref{fig:inos} we show contours of  $\chi_1^\pm$ total decay width $\Gamma_{\chi_1^\pm} = 2$, 5, 20 and 50 keV, in the  $[\mu, M_1]$ plane assuming $M_2=\mu$ and $\tan\beta = 1$. Is it clear that it is possible to simultaneously attain $m_{\chi_1^\pm} =375$ GeV and a very small width, $\Gamma_{\chi_1^\pm} \lesssim 5$ keV, which allows
 a sufficient enhancement of the $\chi_1^\pm$ loop contribution to the $A \chi_1^\pm
\chi_1^\mp$ amplitude to reproduce the diphoton rate measured at the LHC. 
Hence, the situation is not desperate in the MSSM and there is a way to explain the 
properties of the diphoton resonance in this context. 


Let us now turn to the case of a two--Higgs doublet model (2HDM) \cite{2HDM}, again identifying the pseudoscalar state $A$ with the 750 GeV resonance. The phenomenology 
of $A$ is exactly the same as in the MSSM, in particular in the 2HDM alignment limit 
in which the lighter $h$ state is SM--like and assuming the charged $H^\pm$ boson to be heavy enough.  The $A$ production mode at the LHC is the same as discussed above, but for the two--photon decay the only contribution will be that coming from top quark loops which, much like in the MSSM case, is again too small:  an enhancement factor of at least $\approx 400$ is required to obtain the resonance cross section of $\sigma(gg\to A \to \gamma\gamma) \approx 6\pm 2$ fb measured at the LHC. Part of this enhancement can be obtained  by introducing a doublet and two singlets of heavy vector--like leptons\footnote{This is needed in order, first to cancel the chiral anomalies and second, to arrange that the lightest Higgs coupling to two photons, which is measured to be SM--like, is not significantly altered; see Ref.~\cite{ADM}.}  
\beq
L_{L/R}= \left( \!  \begin{array}{c} N  \\  E \end{array}  \! \right)_{ \hspace*{-1mm}L/R\;}, \   
E^{\prime }_{L/R} \ , \ N^{\prime}_{L/R} \ ,  
\label{eq:VLL}
\eeq
with the minimal Lagrangian describing their Yukawa couplings in the interaction basis 
\begin{align} 
- {\cal L}_{\rm Y}  =  \bigg \{ 
     \frac{y_{L}^{E}}{\sqrt{2}}  \overline{L}_L \Phi_dE'_R  
 +   \frac{y_{L}^{N}}{\sqrt{2}}  \overline{L}_L  \Phi_u N'_{R}   
 +  {\rm  L \! \leftrightarrow \! R }  
 +  m_L\overline{L}_L L_R + m_N \overline{N'}_L  N'_R  + m_E \overline{E}'_L E'_R \bigg \}
\!+\!  {\rm h.c.}, 
\label{eq:LagVLQ}
\end{align}
 The Yukawa terms will result in mixing between the doublet and singlets, with the mixing matrix of the neutral/charged leptons taking the form:
\begin{equation}
\mathcal{M}_N=\left(\begin{array}{cc}m_N & \tfrac{1}{\sqrt{2}}y_L^{N} v_{u}\\
\tfrac{1}{\sqrt{2}}y_L^{N} v_{u} & m_{L}\end{array}\right) \ , \qquad   \qquad \mathcal{M}_E=\left(\begin{array}{cc}m_L & \tfrac{1}{\sqrt{2}}y_L^{E} v_{d}\\
\tfrac{1}{\sqrt{2}}y_L^{E} v_{d} & m_{E}\end{array}\right).
\end{equation}
On diagonalizing these matrices with angles $\theta_{N,E}$, the mass eigenstates can be written as
\begin{align}
N_1 & =\cos\theta_N N'+ \sin\theta_N N \, ,  \ N_2 =  \cos\theta_N N- \sin\theta_N N' \,  , \  
\tan2\theta_N=\sqrt{2}y_L^{N} v_{u}/(m_L-m_N) , \nonumber \\
E_1 & = \cos\theta_E E+ \sin\theta_E E' \, , \ \ \  E_2 = \cos\theta_E E'-\sin\theta_E E \, , \  \ 
\tan2\theta_E={\sqrt{2} y_L^{E} v_{d}}/(m_E-m_L) . \nonumber
\end{align}

In light of the data on the diphoton resonance, one then assumes that the lepton $E_1$ has a mass $m_{E_1} \approx 375$ GeV and a Yukawa coupling $y_L^E \approx 2$, a value that is slightly below the perturbative  limit. This allows an initial enhancement of the form factor $A_{1/2}^A$ of the $A\to \gamma \gamma$ amplitude. Nevertheless, to arrive at the needed  value of $\sigma \approx 6$  fb for the diphoton rate, the $E$ amplitude needs to be further enhanced by a factor of about 4 to 6. In the original scenarios, see e.g. Ref.~\cite{ADM}, several replicas of the above spectrum were needed, leading to a model that is far from being minimal. This additional factor can be now generated by the threshold enhancement of $A_{1/2}^A$ as  it was discussed before. 
  
Indeed, if one assumes $m_{E_1}=\frac12 M_A$ within a few MeV and a very small $E_1$ total decay width $\Gamma_{E_1} \ll 1$ MeV, an order of magnitude enhancement of the $A\to \gamma\gamma$ amplitude can be be obtained with the minimal lepton spectrum of eq.~(\ref{eq:VLL}). The small width $\Gamma_{E_1}$ can be obtained simply by ensuring that the mass difference $m_{E_1}-m_{N_1}$ is small and positive (this near mass degeneracy is required anyway in order to comply with precision electroweak data \cite{ADM}). This makes that the only possible $E_1$ decay is the three--body mode $E_1 \to N_1 W^* \to  N_1 f\bar f'$ which requires a highly virtual $W$ boson that strongly suppresses the decay width.

\begin{figure}[btp]
\centerline{
\includegraphics[width=0.41\textwidth]{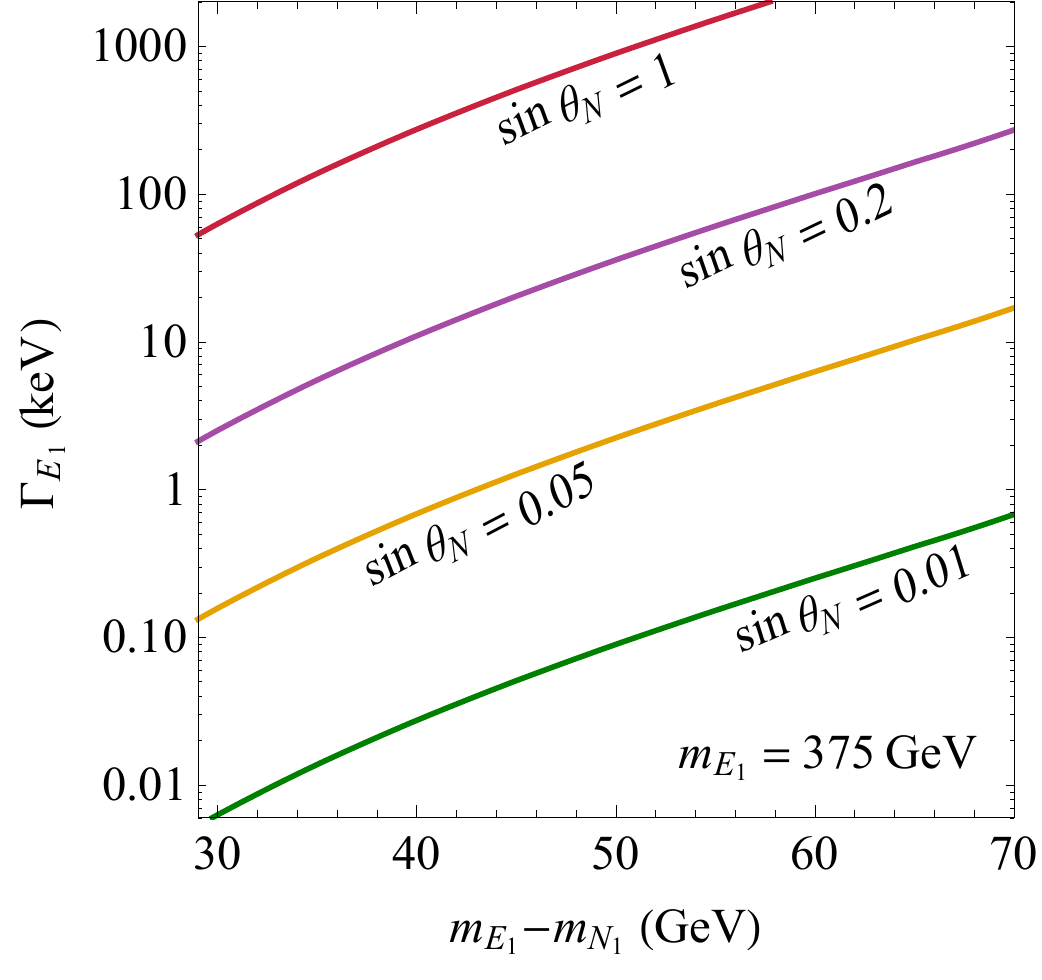}\hspace{.5cm}
\includegraphics[width=0.39\textwidth]{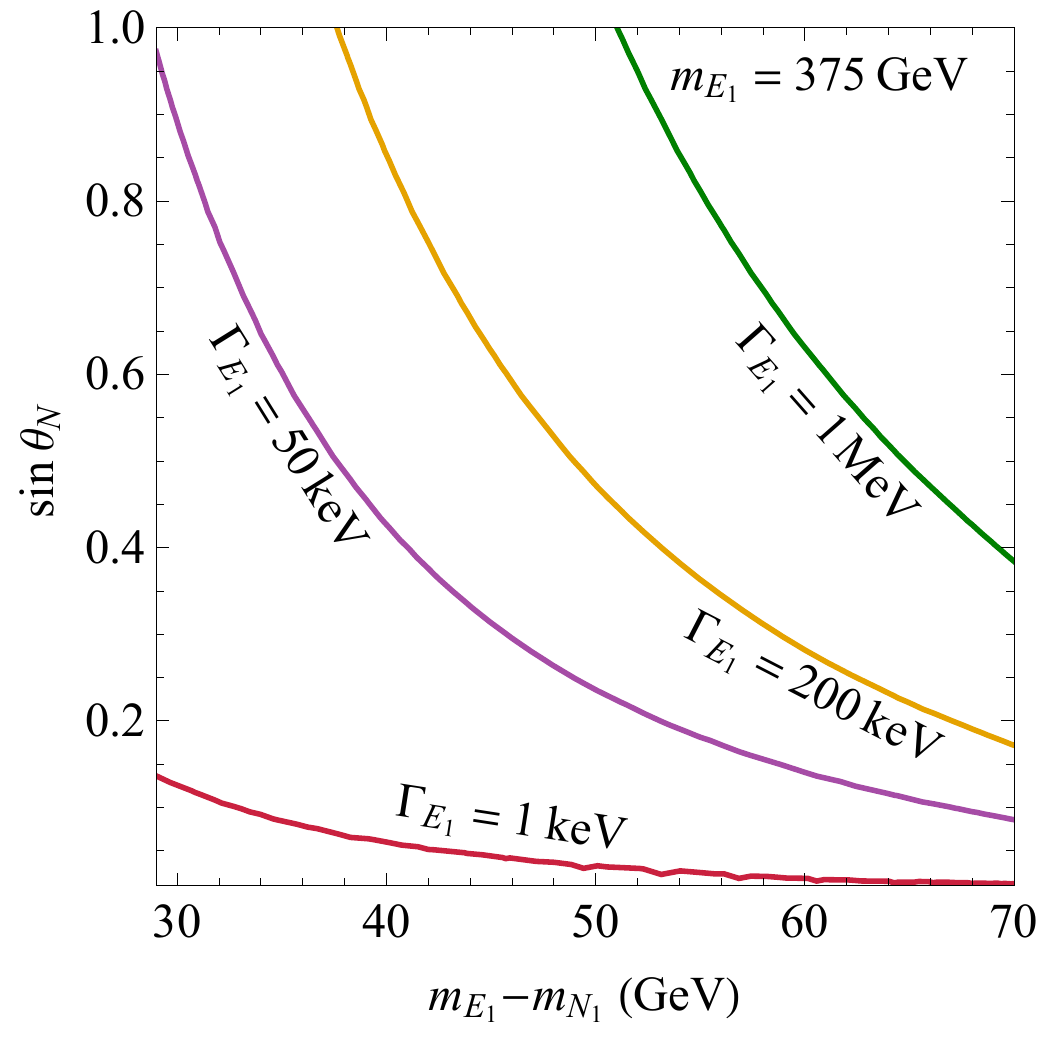}
}
\vspace*{-2mm}
\caption{Left: the sum of partial three--body decay widths $\Gamma (E_1 \to N_1 W^* \to  N_1 f\bar f')$ in GeV as a function of the mass difference $m_{E_1}-m_{N_1}$ for the values of $\sin\theta_N$  as shown. Right: the mixing angle between the doublet and singlet as a function of the mass difference $m_{E_1}-m_{N_1}$ which results in the width $\Gamma_{E_1}$ as indicated. }
\label{fig:plots}
\vspace*{-4mm}
\end{figure}

These decays of the heavy leptons have been discussed in Ref.~\cite{lepton-decay} and using 
the relevant formulae for the three--body  $E_1 \to N_1 W^* \to  N_1 f\bar f'$ channel provided in the papers above, we show in the left--hand side of Fig.~\ref{fig:plots}, the partial decay width (which in the  absence of fermion mixing corresponds to the total width), $\Gamma_{E_1}=\Gamma(E_1 \to N_1 W^* \to  N_1 f\bar f')$, as a function of $m_{E_1}-m_{N_1}$, assuming $\cos\theta_E=1$.. As can be seen a small width of about $\Gamma_{E_1} \approx 1$ keV can be obtained for a 50 GeV mass difference when $\sin\theta_N \simeq0.033$. On the right--hand side of Fig.~\ref{fig:plots}, the mixing angle $\sin\theta_N$ is shown as a function of the mass difference $m_{E_1}-m_{N_1}$ for various values of the width $\Gamma_{E_1}$. From Fig.~\ref{fig:Ratio-Gamma}, we have seen that for widths below $\sim 20$ keV the desired enhancement factor can be obtained. Therefore, a mass difference in the range $m_{E_1}-m_{N_1} \! =\! 40$--80 GeV could easily lead to the 4--6 enhancement factor needed to explain the LHC diphoton data, assuming that the
$A$ resonance is produced via gluon fusion.

Another possibility that is worth considering, although not in the 2HDM context,  is if the $\Phi$ resonance is produced via photon fusion mediated by lepton loops in the same manner as the decay~\cite{photon}. The enhancement factor $F$ would therefore enter both in the production and the decay. Note that, as the requirement of a 45 GeV total width preferred by ATLAS needs not be imposed, the A coupling to two photons
makes that $A\to \gamma\gamma$ may even be the dominant decay mode. In the case of production by photon fusion, the cross section at the $\sqrt{s}=13$ TeV LHC can be expressed as \cite{photon}
\begin{equation}
\label{eq:photonfusion}
\sigma (\gamma\gamma\to A\to\gamma\gamma) \simeq (91-240\,{\rm fb}) \times \Gamma_A^{\rm tot} \, ({\rm GeV}) \times [ {\rm BR}(A\to\gamma\gamma)]^2
\end{equation}
where the range in fb is due to the uncertainty in describing inelastic contributions to the photon parton distribution function. For a given total width $\Gamma_A^{\rm tot}$, eq.~\eqref{eq:photonfusion} can be used to obtain the necessary $A\to \gamma \gamma$ partial width in order to reproduce the observed $\sigma (A)=4$--8 fb inclusive cross section. It is then trivial to calculate the necessary enhancement factor $F$ in order to achieve such a partial width as a function of the total width $\Gamma_A^{\rm tot}$.  One finds that if the dominant contribution to the total width originates from the $\gamma\gamma$ mode, the signal observed at the LHC could be explained assuming photon fusion production if the threshold enhancement factor $F$ is about 2. If, in turn, the total width is closer to $\Gamma_A^{\rm tot}=45$ GeV, an enhancement factor $F$ of about 20  would be required. 

\subsection*{4. Implications for Dark Matter} 

Since in both the scenarios we have studied, the lightest non-SM particles (the lightest neutralino and the  mostly singlet lightest heavy neutrino) are charge and color--neutral, it is tempting to examine their viability as dark matter candidates. In order to simplify the discussion we assume as usual some discrete symmetry (R--parity in the MSSM and a ${\cal{Z}}_2$ symmetry in the 2HDM case under which the heavy leptons are even whereas all other fields are odd) that renders the lightest new state completely stable. This assumption has actually  already, explicitly or implicitly, been made in the previous sections.

We begin with the 2HDM scenario, where the situation turns out to be more straightforward. Focusing on the regime where the dark matter annihilation is mediated by the $s$-channel exchange of the pseudoscalar $A$ state and including all relevant interactions between the dark matter particles and the SM ones, notably those mediated by the $Z$ boson,  the relevant part of the Lagrangian eq.~\eqref{eq:LagVLQ} in terms of mass eigenstates can be written as 
\begin{align}\label{eq:dmlagrangian}
{\cal{L}} = {\cal{L}}_{\rm SM}  - i \frac{y_L^N}{\sqrt{2}} s_{\theta_N} c_{\theta_N} A \bar{N}_1 \gamma^5 N_1 - i \frac{y_t}{\sqrt{2}} A \bar{t} \gamma^5 t - i \frac{y_b}{\sqrt{2}} A \bar{b} \gamma^5 b + \frac{e}{2 s_W c_W} s_{\theta_N}^2 \bar{N}_1 \gamma^\mu N_1 Z_\mu 
\end{align}
where $s_{\theta_N} \equiv \sin\theta_N$, with similar notations in the charged sector. Note that we have ignored $A$ couplings to light fermions since their contributions, being Yukawa suppressed, are much smaller than those of top and bottom quarks as well as those mediated by  the $Z$--boson. We will moreover restrict our analysis to $N_1$ masses below $\sim 350$ GeV. This choice is motivated by the fact that for smaller $E_1, N_1$  mass splittings, co-annihilation processes become important and should be taken into account\footnote{In the co-annihilation region we would have to include a proper treatment of the -- potentially Sommerfeld-enhanced~\cite{Sommerfeld} -- $E_1 E_1$ annihilation, see e.g.~Ref.~\cite{SommerfeldRD}, a task which goes well beyond our purposes.}  \cite{Baker:2015qna}. For simplicity, we thus stick to parameter space regions where these processes are expected to be subleading and can be neglected. To study the dark matter aspects of the model, we have implemented the Lagrangian eq.~\eqref{eq:dmlagrangian} in the public code MicrOMEGAs \cite{Belanger:2014vza} with the help of the FeynRules package \cite{Alloul:2013bka}.

\begin{figure}[t!]
\centerline{ \includegraphics[width=0.65\textwidth]{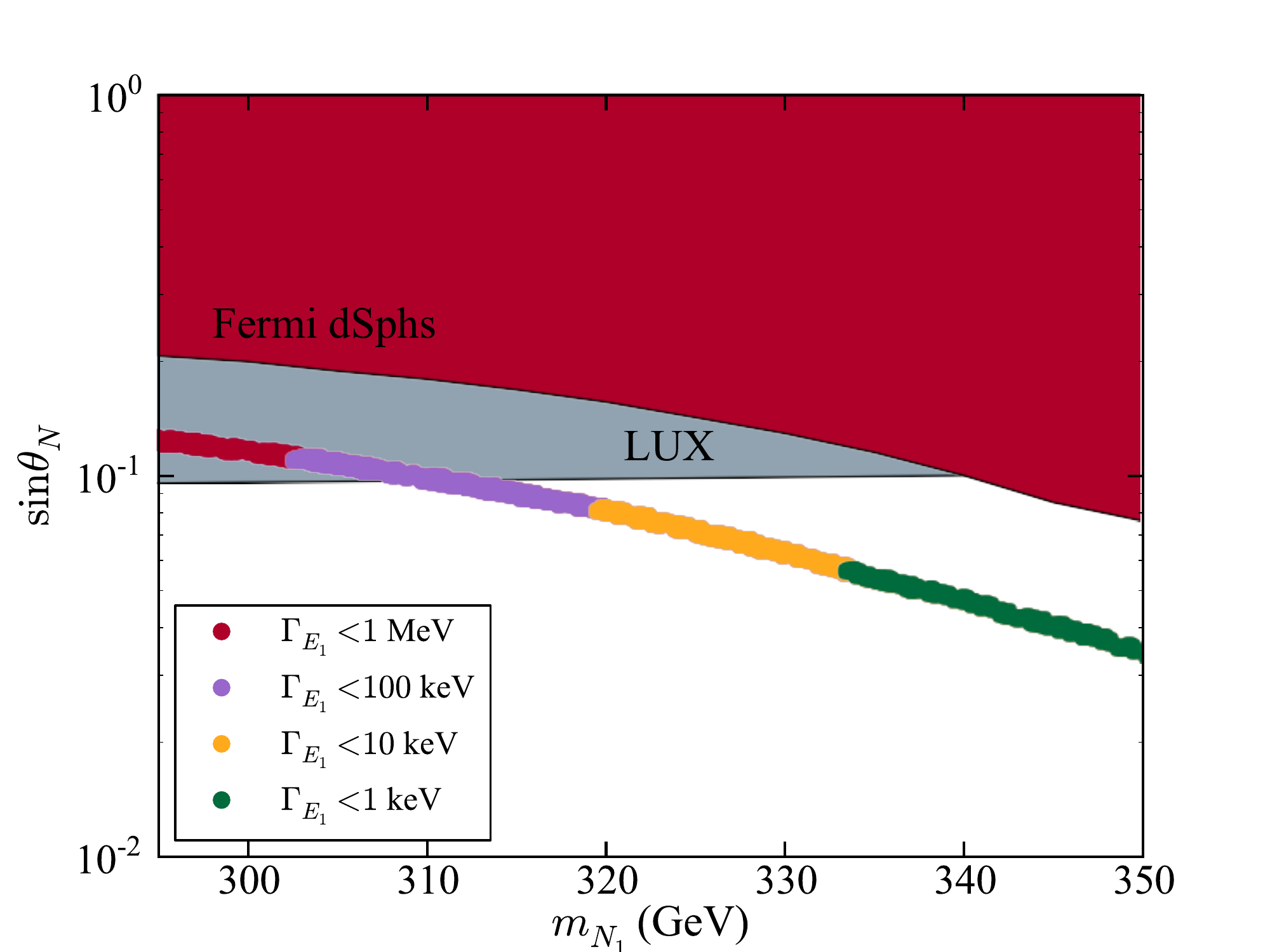} }
\caption{Mixing angle versus the mass of the lightest heavy neutrino for which the Planck bound on the dark matter abundance in the universe is satisfied (colored band) for different widths of the lightest heavy electron, $\Gamma_{E_1}$. In the grey shaded region, the predicted spin-independent scattering cross-section off nucleons is in conflict with the latest LUX limits. The red shaded region depicts the limits from the Fermi satellite searches for dark matter annihilation -- induced continuum gamma rays in dSphs.
}
\label{fig:dm}
\end{figure}

Our results are depicted in Fig.~\ref{fig:dm}, where we highlight the $(m_{N_1}, \sin\theta_N)$ combinations for which the latest limits on dark matter abundance from the Planck mission \cite{Ade:2015xua} can be satisfied according to standard thermal freeze-out. The different colorings correspond to different ranges for the predicted width of the lightest heavy electron $E_1$ from the decay mode $E_1 \rightarrow N_1 W^{*} \to N_1 f\bar f'$. The other parameters entering the Lagrangian eq.~\eqref{eq:dmlagrangian} have been set to the values $y_L^N =y_t = 1$ (which also amounts to a total width $\Gamma_A \approx 40$ GeV), while for the calculation of the $E_1$ decay width we have chosen $m_{E_1} = 375.003$ GeV and $s_{\theta_E} = 0.1$. The heavier neutrino mass $m_{N_2}$ has been set to a large value in this analysis. 

A well-known constraint on dark matter scenarios involving vector--like couplings to the $Z$ boson comes from direct detection; see for example Ref.~\cite{Arcadi:2013qia}. We have computed the predicted spin-independent scattering cross section off nucleons\footnote{The spin-dependent scattering is found to be much weaker and will be ignored. Besides, we recall that pseudoscalar couplings of Dirac dark matter to the SM particles yield a negligible spin-independent scattering cross section, the latter being proportional to the momentum transfer which is extremely small compared to the mass $M_A$.} and compared it to the updated analysis performed by the LUX collaboration \cite{Akerib:2015rjg}. The excluded regions of parameter space are depicted by the gray-shaded area of Fig.~\ref{fig:dm}.  As expected, the direct detection limits restrict the mixing of the singlet and doublet heavy neutrinos to small values, where the annihilation is mostly dominated by $A$--boson exchange. For our choice of parameters, this constraint also forces small values for the total width of the heavy electron $\Gamma_{E_1}$. 

Additional constraints on the scenario come from searches for dark matter annihilation -- induced gamma rays and in particular from the Fermi satellite searches for continuum gamma rays in Dwarf Spheroidal Galaxies (dSphs) \cite{Ackermann:2015zua} and for gamma ray lines from the Galactic center \cite{Ackermann:2015lka}. The former are depicted by the red band in Fig.~\ref{fig:dm}. As for the latter, we find them to be subdominant throughout our parameter space even assuming a realistic halo profile for our Galaxy \cite{NFW-Einasto}. This might appear to be slightly counter-intuitive, since we are invoking here a mechanism that significantly boosts the diphoton signal. However, the threshold enhancement which is effective in the LHC environment is actually irrelevant in the case of indirect detection, due to the fact that the center of mass energy is not sufficient to produce the mediator $A$ on--shell. This is actually an attractive up-shot of the threshold enhancement mechanism invoked in our work, which relieves the tension that has been shown to exist \cite{D'Eramo:2016mgv} in models attempting to relate the 750 GeV diphoton excess with dark matter.

All in all, we see that the relic density, direct and indirect detection constraints are compatible with the small $E_1$ width values needed in order to reproduce the LHC diphoton excess, yielding a viable dark matter candidate under the form of a mostly singlet heavy neutrino. Moreover, at least for the parameter ranges depicted in Fig.~\ref{fig:dm}, we expect that direct and indirect detection experiments will be able to probe the Planck compatible parameter space region within the next few years.

We now turn to the MSSM case and for our computations, we again employ MicrOMEGAs. We fix, as discussed previously, the lightest chargino mass at $m_{\chi_1^\pm} =\frac12 M_A =375$ GeV and require a mass difference $m_{\chi_1^\pm} - m_{\chi_1^0} < M_W$ to ensure a small decay width for
$\chi_1^\pm$. The lightest neutralino turns out to be an admixture of bino, higgsino and wino, with an under-abundant relic density $\Omega h^2 \sim 10^{-3}$ -- $10^{-2}$, as a result on one hand of the relatively strong couplings of mixed neutralino scenarios to $A$, and on the other hand of the neutralino mass being relatively  close to $\frac 12 M_A$, i.e. the so--called ``funnel region''. Thus, in this scenario, thermal relic neutralinos cannot account for more than ${\cal{O}}(10\%)$ of the total dark matter in the universe. Note that even for under-abundant dark matter components direct detection bounds do apply, upon appropriate rescaling of the limits. We find that in the relevant region of parameter space, the combination $\sigma_{\rm SI} \times \Omega_{\rm MSSM}/\Omega_{\rm Planck}$ lies below the LUX bounds. 

In light of these findings, an interesting possibility would be to consider the option of gravitino dark matter, potentially upon embedding of our MSSM scenario in a (most likely general \cite{Meade:2008wd}) gauge-mediated supersymmetry breaking framework. Given that the gravitino abundance from neutralino decays will in general decrease as $m_{\tilde{G}}/m_{\chi_1^0}$ with respect to the -- already under-abundant -- neutralino relic density, the most likely scenario would in fact be thermal gravitino production; see e.g. Ref.~\cite{Rychkov:2007uq} and references therein. Performing such an analysis goes beyond the scope of our study. In any case, gravitino dark matter with a general neutralino next-to-lightest superparticle has been extensively studied in Ref.~\cite{Covi:2009bk}. Moreover, the under-abundance of these neutralinos should help relax the tension with Big Bang Nucleosynthesis constraints, see e.g. the recent discussion in Ref.~\cite{Arvey:2015nra}.

\subsection*{5. Conclusions} 

In this paper, we have discussed the possibility of threshold enhancing the branching ratio of the decay into two photons for a 750 GeV pseudoscalar boson $A$, in light of the recent experimental hints of an excess in the diphoton  spectrum at the LHC. If the loop mediating the $A \rightarrow \gamma \gamma$ decay was to contain new fermions at approximately half the mass of the resonance, i.e.~$\frac12 M_A\! \sim \! 375$ GeV, then this decay could be significantly enhanced. The precise value of the enhancement factor was shown to depend on the width and the mass of these new fermions. Concretely, we found that fermion widths smaller than $\Gamma_f<100$ keV, naturally occurring in 3-body decay processes, and masses $m_f$ of a few MeV above 375 GeV could lead to enhancement factors of 2 to 20 at the amplitude level. We then applied this idea to two concrete new physics scenarios where $A$ could be produced via the gluon fusion mechanism: the minimal supersymmetric model and a two-Higgs doublet model augmented with one vector-like doublet and two singlets of leptons.

In the MSSM case, we found that a chargino with mass $m_{\chi_1^\pm} \approx 375$ GeV can provide the necessary enhancement factor to attain a diphoton cross section of the order of 6 fb as favoured by ATLAS and CMS. We examined the chargino total decay width, where the chargino decays into a neutralino and a SM fermion pair through a sufficiently off-shell $W$ boson. If this width lies below $\sim 2$ keV, a condition which can easily be satisfied for appropriate choices of the soft masses $M_{1,2}$ and the higgsino mass parameter $\mu$, the necessary enhancement in the chargino loop in order to reproduce the observed diphoton excess, of the order of 20 at the amplitude level, is obtained. To the best of our knowledge, this is the only explanation of the 750 GeV diphoton excess that has been proposed in the literature within the plain MSSM without any additional particle content (for an extension like the NMSSM with no additional particles, see for instance Ref.~\cite{NMSSM}).

We then discussed the threshold enhancement mechanism in the context of a basic 2HDM in which  vector-like leptons are added to the spectrum. The lightest (mostly part of an isodoublet which also contains a neutral lepton) charged lepton $E_1$ is responsible for the threshold enhancement and its width is again given by the three-body decay $E_1 \! \to \! N_1\bar{f}f'$
(with $N_1$ being mostly an isosinglet).  In a large region of parameter space, it is found to lie in the desired region $\Gamma_{E_1}  < 50$ keV, resulting in an enhancement factor of the order of 4--6 at the amplitude level as required to explain the LHC diphoton data.

We moreover briefly commented upon the possibility that the diphoton resonance is produced at the LHC via photon fusion, $\gamma\gamma \to A \to \gamma\gamma$. In this case, the necessary value of the enhancement factor ranged from about 2 to 20, where the limiting values represent the cases where the decay of the $A$ state is restricted to the $\gamma\gamma$ mode or where the total width is taken to be of the order of 45 GeV, as favoured by the ATLAS data.

As a final exercise we studied whether the lightest neutral states of the new physics spectrum, the lightest neutralino in the MSSM and the lightest vector-like neutrino in our 2HDM variant, can play the role of dark matter in the Universe, where appropriate symmetries prevent the decay of these into SM particles. In our 2HDM scenario, we found that for the mass range from $m_{N_1} \approx 315$ to 350 GeV it is perfectly possible to satisfy the Planck constraints on the dark matter density in the Universe while being compatible with the LUX limits on the spin-independent scattering cross section off nuclei and the Fermi-LAT indirect searches for continuum $\gamma$-rays from dark matter annihilation in Dwarf Spheroidal Galaxies. Searches for gamma-ray lines were found to provide subleading constraints, since the threshold enhancement mechanism is not effective in dark matter annihilation at low velocities. In the MSSM case, the neutralino relic density turns out to be below the Planck value such that the direct detection constraints do not affect the region of interest.

Finally, let us note again that the scenarios exhibiting threshold enhanced diphoton signals are extremely contrived as they only occur in very narrow ranges of parameter space, i.e. of $m_f\! -\! \frac12 M_A$; therefore fine tuning at the 10 to 100 keV is required which may appear unnatural. Nevertheless, as it allows one to avoid complicated scenarios (with possibly a large multiplicity of new fermions) that are sometimes at the verge of being non-perturbative, Occam’s razor leads us to believe that this “fine-tuned” scenario could constitute a plausible option.
\bigskip

\noindent {\bf Acknowledgements:}  We thank the CERN Theory Department for its hospitality
during the completion of this work as well as Manuel Drees and Pedro Ruiz-Femenia for a  careful reading of the manuscript and helpful comments. This work is supported by the ERC advanced grant Higgs@LHC. A.G. is supported by the `New Frontier's" program of the Austrian Academy of Sciences.


\begin{thebibliography}{999}
\begin{small}

\bibitem{diphoton} ATLAS Collaboration, ATLAS-CONF-2015-081; CMS Collaboration, CMS-PAS EXO-15-004.

\bibitem{cite-few} See e.g., R.~Franceschini et al., arXiv:1512.04933 [hep-ph];
J. Ellis, S.A.R. Ellis, J. Quevillon, V. Sanz and  T. You, arXiv:1512.05327 [hep-ph]; 
M.~R.~Buckley, arXiv:1601.04751 [hep-ph]; A. Djouadi, J. Ellis, R. Godbole and J. Quevillon,  arXiv:1601.03696 [hep-ph],  F.~Staub {\it et al.},  arXiv:1602.05581 [hep-ph].

\bibitem{ADM} A. Angelescu, A. Djouadi and G. Moreau, arXiv:1512.04921 [hep-ph].  

\bibitem{Reviews} A.~Djouadi, Phys. Rept. 459 (2008) 1.

\bibitem{PT} See for instance,
M. Dhuria and G. Goswami, arXiv:1512.06782 [hep-ph]; 
F. Goertz, J.F. Kamenik, A. Katz and M. Nardecchia, arXiv:1512.08500 [hep-ph]; 
M. Fabbrichesi and A. Urbano,  arXiv:1601.02447 [hep-ph]; 
E. Bertuzzo, P. Machado and M. Taoso, arXiv:1601.07508 [hep-ph]; 
A.~Salvio, F.~Staub, A.~Strumia and A.~Urbano,  arXiv:1602.01460 [hep-ph].

\bibitem{venerable} J. Ellis, M. Gaillard and D. Nanopoulos,  Nucl. Phys. B106
(1976) 292; H. Georgi, S. Glashow, M. Machacek and D. Nanopoulos, Phys. Rev. Lett. 40 (1978) 692; A.I. Va\u\i nshte\u\i n, M. Voloshin, V. Zakharov and M. Shifman, Sov.\ J. Nucl.\ Phys.\ 30 (1979) 711; J. Gunion, H. Haber, G. Kane and S. Dawson, ``The Higgs Hunter's
Guide", Reading 1990; A.~Djouadi,  Phys. Rept. 457 (2008) 1.

\bibitem{singular} See for instance, M.~Drees and K.~i.~Hikasa, Phys.\ Rev.\ D41 (1990) 1547. 

\bibitem{MSY} K. Melnikov, M. Spira and O. Yakovlev, Z. Phys. C64 (1994) 401.  

\bibitem{GreensFn} V.S.~Fadin and V.~Khoze, Sov.\ J.\ Nucl.\ Phys.\  {48} (1988) 309
[Yad.\ Fiz.\  {48} (1988) 487], JETP Lett.\  {46} (1987) 525   [Pisma Zh.\ Eksp.\ Teor.\ Fiz.\  {46} (1987) 417].

\bibitem{2HDM}  For a review on 2HDMs, see G. Branco et al.,  Phys. Rept. 516 (2012) 1. 

\bibitem{photon} See for instance, 
S. Fichet, G. von Gersdorff and C. Royon, arXiv:1512.05751 [hep-ph]; arXiv:1601.01712 [hep-ph];
C. Csaki, J. Hubitz and J. Terning, arXiv:1512.05776 [hep-ph]; 
U. Danielsson et al., arXiv:1601.00624 [hep-ph];  
H. Ito, T. Moroi and Y. Takaesu, arXiv:1601.01144 [hep-ph]; 
L. Harland-Lang, V. Khoze and M. Ryskin, arXiv:1601.07187 [hep-ph]; 
 arXiv:1601.03772 [hep-ph];  S.~Abel and V.~V.~Khoze, arXiv:1601.07167 [hep-ph]. 

\bibitem{DM} See e.g. Y. Mambrini, G. Arcadi and A. Djouadi, Phys. Lett. B755 (2016) 426  [arXiv:1512.04913]; M. Backovic, A. Mariotti and D. Redigolo, arXiv:1512.04917 [hep-ph]; 
D. Barducci, A. Goudelis, S. Kulkarni and D. Sengupta, arXiv:1512.06842 [hep-ph];
P.B.~Dev and D.~Teresi, arXiv:1512.07243 [hep-ph];
P. Ko and  T. Nomura, arXiv:1601.02490 [hep-ph].  

\bibitem{LHC-VLQ} ATLAS Collaboration, JHEP 08 (2015) 105, arXiv:1503.05425, arXiv:1505.04306, arXiv:1509.04261; CMS Collaboration, Phys. Lett. B729 (2014) 149.

\bibitem{Q-boundstates} M.x.~Luo et al., arXiv:1512.06670 [hep-ph]; 
C. Hand et al.,  arXiv:1602.08100 [hep-ph]; 
Y.~Kats and M.~Strassler, arXiv:1602.08819 [hep-ph].  

\bibitem{SDGZ} M. Spira, A. Djouadi, D. Graudenz and P.M. Zerwas,  Phys. Lett. B318 (1993) 347; Nucl. Phys. B453 (1995) 17.  

\bibitem{Htt} ATLAS collaboration, JHEP 1508 (2015) 148; CMS collaboration, arXiv:1506.03062. 

\bibitem{Coulomb} For a recent discussion, see for instance M. Beneke et al., arXiv:1601.04718 
[hep-ph]. 

\bibitem{hMSSM} These numbers have been obtained in the context of the hMSSM discussed in
A. Djouadi et al., JHEP 1506 (2015) 168; JHEP 1310 (2013) 028; Eur. Phys. J. C73 (2013) 2650.

\bibitem{hdecay}  The numerical analysis is performed using the program {\tt HDECAY} and 
{\tt FeynHiggs}: A.~Djouadi, J.~Kalinowski and M.~Spira, Comput. Phys. Commun. 108 (1998) 56; A. Djouadi, M. Muhlleitner and M. Spira, Acta. Phys. Polon. B38 (2007) 635; S. Heinemeyer, W. Hollik and G. Weiglein, Comput. Phys. Commun. 124 (2000) 76. 

\bibitem{lepton-decay} A. Djouadi, Z. Phys. C63 (1994) 317; G. Azuelos and A. Djouadi, Z. Phys. C63 (1994) 327.


\bibitem{Sommerfeld} A. Sommerfeld, Ann. Phys. 11 (1931) 257

\bibitem{SommerfeldRD}
  J.~Hisano, S.~Matsumoto, M.~Nagai, O.~Saito and M.~Senami,
  Phys.\ Lett.\ B {\bf 646} (2007) 34.

\bibitem{Baker:2015qna} M.~J.~Baker et al., JHEP 1512 (2015)  120. 

\bibitem{Belanger:2014vza} G.~B\'elanger, F.~Boudjema, A.~Pukhov and A.~Semenov,
  Comput.\ Phys.\ Commun.\  {92} (2015) 322. 

\bibitem{Alloul:2013bka} 
  A.~Alloul, N.~D.~Christensen, C.~Degrande, C.~Duhr and B.~Fuks,
  Comput.\ Phys.\ Commun.\  185  (2014) 2250.  

\bibitem{Ade:2015xua} 
  P.~A.~R.~Ade {et al.} [Planck Collaboration],
  arXiv:1502.01589 [astro-ph.CO].

\bibitem{Arcadi:2013qia} 
  G.~Arcadi, Y.~Mambrini, M.~H.~G.~Tytgat and B.~Zaldivar,
  JHEP 1403 (2014) 134.  

\bibitem{Akerib:2015rjg} 
  D.~S.~Akerib {et al.} [LUX Collaboration],
  arXiv:1512.03506 [astro-ph.CO].

\bibitem{Ackermann:2015zua} 
  M.~Ackermann et al. [Fermi-LAT Collaboration],
  Phys.\ Rev.\ Lett.\  115, no. 23 (2015) 231301. 

\bibitem{Ackermann:2015lka} 
  M.~Ackermann et al. [Fermi-LAT Collaboration],
  Phys.\ Rev.\ D 91 no. 12 (2015) 122002. 

\bibitem{D'Eramo:2016mgv} 
  F.~D'Eramo, J.~de Vries and P.~Panci,
  arXiv:1601.01571 [hep-ph].

\bibitem{NFW-Einasto} J.~F.~Navarro, C.~S.~Frenk and S.~D.~M.~White, Astrophys.\ J.\ 490 (1997) 493; A.~W.~Graham, D.~Merritt, B.~Moore, J.~Diemand and B.~Terzic, Astron.\ J.\ {132} (2006)  2685.

\bibitem{Meade:2008wd} 
  P.~Meade, N.~Seiberg and D.~Shih,
  Prog.\ Theor.\ Phys.\ Suppl.\   177 (2009) 143. 

\bibitem{Rychkov:2007uq} 
  V.~S.~Rychkov and A.~Strumia,
  Phys.\ Rev.\ D75 (2007) 075011.

\bibitem{Covi:2009bk} 
  L.~Covi, J.~Hasenkamp, S.~Pokorski and J.~Roberts,
  JHEP 0911 (2009) 003. 

\bibitem{Arvey:2015nra} For a recent discussion, see for instance:  
  A.~Arbey, M.~Battaglia, L.~Covi, J.~Hasenkamp and F.~Mahmoudi,
  Phys.\ Rev.\ D92, no. 11 (2015) 115008. 
%

\bibitem{NMSSM} U. Ellwanger and C. Hugonie, arXiv:1602.03344 [hep-ph]; 
F. Domingo, S. Heinemeyer, J.S. Kim and K. Rolbiecki, arXiv:1602.07691 [hep-ph]; 
M.~Badziak, M.~Olechowski, S.~Pokorski and K.~Sakurai, arXiv:1603.02203 [hep-ph].
\end{small}
\end{thebibliography}
\end{document}